\documentclass[sigconf,screen,authorversion]{acmart}

\AtBeginDocument{%
  \providecommand\BibTeX{{%
    \normalfont B\kern-0.5em{\scshape i\kern-0.25em b}\kern-0.8em\TeX}}}

\copyrightyear{2024}
\acmYear{2024}
\setcopyright{acmlicensed}\acmConference[UIST '24]{The 37th Annual ACM Symposium on User Interface Software and Technology}{October 13--16, 2024}{Pittsburgh, PA, USA}
\acmBooktitle{The 37th Annual ACM Symposium on User Interface Software and Technology (UIST '24), October 13--16, 2024, Pittsburgh, PA, USA}
\acmDOI{10.1145/3654777.3676374}
\acmISBN{979-8-4007-0628-8/24/10}

\acmConference[UIST '24]{The ACM Symposium on User Interface Software and Technology}{Oct 13--16,
  2024}{Pittsburgh, PA}

\usepackage{graphicx}
\usepackage{balance}
\usepackage{calc}                       
\usepackage{xspace}         
\usepackage{multirow}                   
\usepackage{enumitem}                   
\usepackage{listings}
\usepackage{fancybox}
\usepackage{soul}
\usepackage{xcolor}

\AtBeginDocument{

}

\setlist{noitemsep,parsep=0pt,partopsep=0pt, leftmargin=10pt} 

\definecolor{lightpink}{RGB}{237,157,202}
\definecolor{lightred}{RGB}{210,121,121}
\definecolor{lightorange}{RGB}{230,170,50}
\definecolor{lightgold}{RGB}{210,194,121}
\definecolor{lightgreen}{RGB}{121,210,121}
\definecolor{lightaqua}{RGB}{121,206,210}
\definecolor{lightblue}{RGB}{121,124,210}
\definecolor{lightpurple}{RGB}{153,102,255}
\definecolor{red}{RGB}{178,34,34}
\definecolor{gray}{RGB}{166,166,166}

\newlength{\dpcircle}
\newlength{\rcircle}
\newlength{\dcircle}

\newcommand{\ie}{{i.e.,}\xspace}
\newcommand{\eg}{{e.g.,}\xspace}
\newcommand{\ea}{{et~al.}\xspace}

\newcommand{\etc}{{etc.}\xspace}
\newcommand{\ADA}{ChatGPT\xspace}

\newcommand{\system}{WaitGPT\xspace} 
\newcommand{\baseline}{Baseline\xspace} 
\newcommand{\user}{Zoey\xspace}

\newcommand{\cut}[1]{\textcolor{red}{}}

\newcommand{\iq}[1]{``{\it{#1}}''}

\newcommand{\rev}[1]{\textcolor{black}{#1}}

\newlength\myheight
\newlength\mydepth
\settototalheight\myheight{Xygp}
\definecolor{codegreen}{rgb}{0,0.6,0}
\definecolor{codegray}{rgb}{0.5,0.5,0.5}
\definecolor{codepurple}{rgb}{0.58,0,0.82}
\definecolor{backcolour}{rgb}{0.95,0.95,0.92}

\lstdefinestyle{mystyle}{
    backgroundcolor=\color{backcolour},
    commentstyle=\color{codegreen},
    keywordstyle=\color{magenta},
    numberstyle=\tiny\color{codegray},
    stringstyle=\color{codepurple},
    basicstyle=\ttfamily\footnotesize,
    breakatwhitespace=false,
    breaklines=true,
    captionpos=b,
    keepspaces=true,
    numbers=left,
    numbersep=5pt,
    showspaces=false,
    showstringspaces=false,
    showtabs=false,
    tabsize=2
}

\definecolor{codebg}{rgb}{0.96,0.96,0.96} 
\definecolor{codeborder}{rgb}{0.85,0.85,0.85} 

\lstdefinestyle{prettyinline}{
  backgroundcolor=\color{codebg},     
  basicstyle=\ttfamily\small,         
  breakatwhitespace=false,            
  breaklines=true,                    
  captionpos=b,                       
  commentstyle=\color{gray},          
  keywordstyle=\color{blue},          
  stringstyle=\color{teal},           
  numbers=none,                       
  numberstyle=\tiny\color{codeborder},
  rulecolor=\color{codeborder},       
  frame=single,                       
  framexleftmargin=2pt,               
  framexrightmargin=2pt,              
  framextopmargin=2pt,                
  framexbottommargin=2pt,             
  framesep=2pt,                       
  xleftmargin=3pt,                    
  xrightmargin=3pt,                   
  aboveskip=5pt,                      
  belowskip=3pt,                      
  showstringspaces=false,             
  upquote=true,                       
}

\newcommand{\op}[1]{%
    \begingroup
  \setlength{\fboxsep}{0pt}%
  \colorbox{gray!15}{%
    \raisebox{0pt}[0.6\height][0.3\height]{%
      \hspace{0.1em}
      \textcolor{black}{\textsc{#1}}%
      \hspace{0.1em}
    }%
  }%
  \endgroup
}

\begin{document}

\title{WaitGPT: Monitoring and Steering Conversational LLM Agent in Data Analysis with On-the-Fly Code Visualization}

\author{Liwenhan Xie}
\authornote{This study was partially conducted during an academic visit to Harvard University.}
\email{liwenhan.xie@connect.ust.hk}
\orcid{0000-0002-2601-6313}
\affiliation{%
  \institution{Hong Kong University of Science and Technology}
  \city{Hong Kong SAR}
  \country{China}
}

\author{Chengbo Zheng}
\email{cb.zheng@connect.ust.hk}
\orcid{0000-0003-0226-9399}
\affiliation{%
  \institution{Hong Kong University of Science and Technology}
  \city{Hong Kong SAR}
  \country{China}
}

\author{Haijun Xia}
\email{haijunxia@ucsd.edu}
\orcid{0000-0002-9425-0881}
\affiliation{%
  \institution{University of California San Diego}
  \city{La Jolla}
  \state{CA}
  \country{USA}
}

\author{Huamin Qu}
\email{huamin@ust.hk}
\orcid{0000-0002-3344-9694}
\affiliation{%
  \institution{Hong Kong University of Science and Technology}
  \city{Hong Kong SAR}
  \country{China}
}

\author{Chen Zhu-Tian}
\email{ztchen@umn.edu}
\orcid{0000-0002-2313-0612}
\affiliation{%
  \institution{University of Minnesota}
  \city{Minneapolis}
  \state{MN}
  \country{USA}}

\renewcommand{\shortauthors}{L. Xie, C. Zheng, H. Xia, H. Qu, and C. Zhu-Tian}

\begin{abstract}
 Large language models (LLMs) support data analysis through conversational user interfaces, as exemplified in OpenAI's ChatGPT (formally known as Advanced Data Analysis or Code Interpreter).
Essentially, LLMs produce code for accomplishing diverse analysis tasks.
However, presenting raw code can obscure the logic and hinder user verification.
To empower users with enhanced comprehension and augmented control over analysis conducted by LLMs, we propose a novel approach to transform LLM-generated code into an interactive visual representation.
In the approach, users are provided with a clear, step-by-step visualization of the LLM-generated code in real time, allowing them to understand, verify, and modify individual data operations in the analysis.
Our design decisions are informed by a formative study (N=8) probing into user practice and challenges.
We further developed a prototype named \system and conducted a user study (N=12) to evaluate its usability and effectiveness.
The findings from the user study reveal that \system facilitates monitoring and steering of data analysis performed by LLMs, enabling participants to enhance error detection and increase their overall confidence in the results.

\end{abstract}

\ccsdesc[500]{Human-centered computing~Natural language interfaces}
\ccsdesc[300]{Human-centered computing~Graphical user interfaces}
\ccsdesc[300]{Human-centered computing~Information visualization}

\keywords{Conversational Data Analysis, LLM Agent, Human-AI Interaction, Generative AI, Code Verification, Visual Programming}

\begin{teaserfigure}
\centering
  \includegraphics[width=0.92\textwidth]{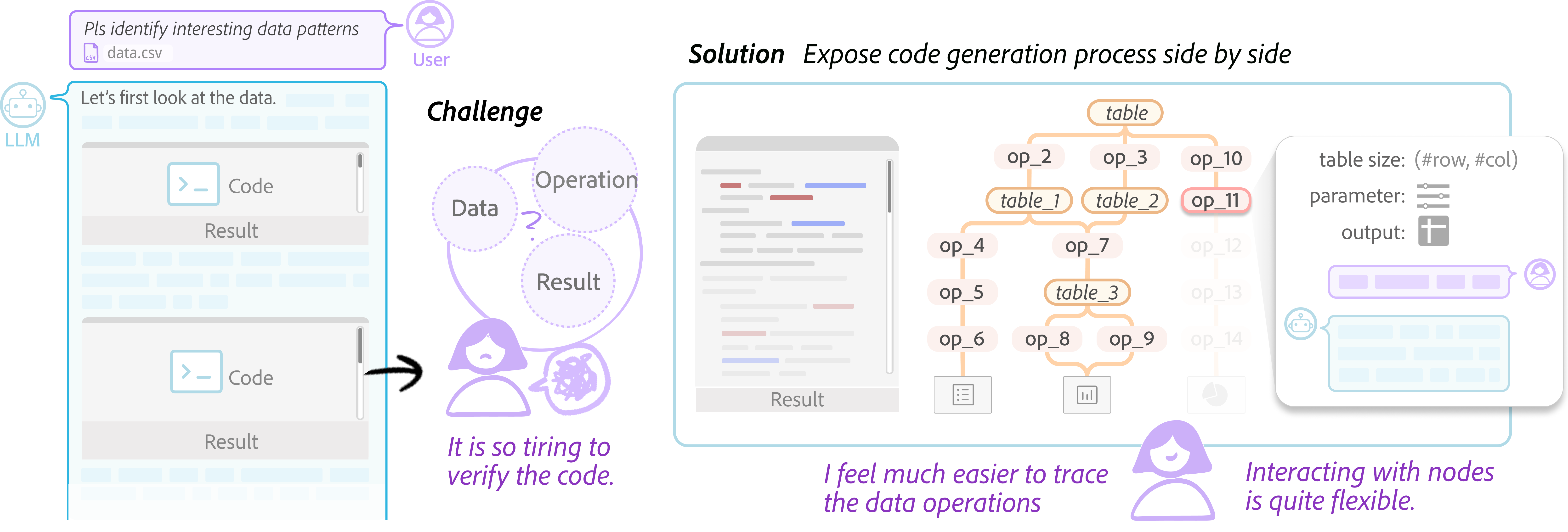}
  \caption{Monitoring and steering LLM-powered data analysis tools with WaitGPT: Beyond viewing the raw code, users can inspect data operations with a transformable representation generated on the fly and participate in data analysis proactively.}
  \Description{Figure 1 compares the traditional conversational interface and WaitGPT's code visualization for LLM-powered data analysis.
Left side: Traditional interface showing challenges users face in tracking data, operations, and results when verifying LLM responses.
Right side: WaitGPT's solution displays a node-link diagram generated in real time alongside the LLM response. The diagram visualizes tables, operations, and results. An expandable box on an operation node allows users to access table size, parameters, and outputs and ask contextual questions.
}
  \label{fig:teaser}
\end{teaserfigure}


\maketitle

\section{Introduction}
Large language models (LLMs) have significantly lowered the entry point for data analysis, empowering users without strong programming skills to engage in sophisticated analytical tasks~\cite{cheng2023gpt, he2023text2analysis, dibia2023lida}. 
Instead of writing scripts or using complex software, people can directly talk to conversational LLM agents.
Examples of emerging LLM-powered data analysis services or tools include ChatGPT Plus~\cite{openai2024ada}, Gemini Advanced~\cite{google2024gemini}, and CodeActAgent~\cite{wang2024codeact}.
Generally, these tools follow a planning framework, where the LLM agent proposes a plan to divide the task, then generates code to process data and continues the process based on the execution result.

Despite their potential, real-world deployment of LLM-powered data analysis tools has exposed reliability concerns, including hallucinations~\cite{liu2023what, chen2024viseval}, subtle bugs~\cite{yang2021subtle, wu2024automated}, and mismatch between LLM's understanding of the tasks and under-articulated user intents~\cite{wang2018mismatch, li2024dawn}.
Such shortcomings necessitate human oversight to verify and correct the data analysis process~\cite{chopra2023conversational, gu2024how, olausson2024self}. 
Current tools often present raw data analysis code, shifting the user's focus to low-level details instead of the high-level data analysis process.
According to our interview with \ADA users, individuals, especially those with limited coding skills, struggle to comprehensively review the code produced by LLMs, 
thereby risking undetected errors and potentially incorrect results.
Moreover, rectifying code through conversation can turn into a cumbersome exchange, adding to the inefficiency and frustration.

\rev{Our goal is to make the data analysis process conducted by LLMs easier to understand and navigate for users, in line with current research on designing UIs featuring generative AIs (\eg~\cite{subramonyam2024bridging, shen2024towards}).
Specifically, we aim to support real-time monitoring and proactive intervention (steering) at any point.
Compared with existing approaches targeting a traditional data analysis pipeline (\eg~\cite{lau2023teaching, shrestha2021unravel}), this scenario features conversational interaction and on-demand generation of unfamiliar code to the users, where the code streams in.
}
Informed by a formative study involving 8 users experienced in LLM-powered data analysis,
we propose a workflow that identifies data operations within the generated code and maps them to visual, interactive primitives on the fly (\autoref{fig:teaser}). 
These primitives collectively offer an overview of the data analysis process, 
and surface the details of each data operation and their internal runtime states in an intuitive, syntax-independent format.
Furthermore, users can refine each operation by interacting directly with these primitives without regenerating the entire analysis code. 
Through this approach, we augment traditional conversational user interfaces (CUIs) with interactive visualization, transforming users from passive recipients of information into active participants in the data analysis task.

We have designed and implemented \system, a prototype system that converts the data analysis code generated by an LLM into a visual diagram that consists of nodes representing key data operations, 
composing an overview step by step.
This diagram progressively evolves along with the code generation process.
Furthermore, \system executes the underlying code line by line and updates the visual diagram to reflect the code's intermediate state during runtime. 
Users can interact with these nodes to modify or adjust the operations, 
thereby refining the data analysis process. 
Execution results are maintained and preserved within a sandbox environment, 
enabling the system to resume or rerun the analysis code after modifications, 
without the need to regenerate the entire code.
A user study with 12 participants reported an enhanced experience, noting the ease of spotting errors, increased agency, and heightened confidence in the results produced by the LLM.

In summary, our contributions are three-fold.
\begin{itemize}
      \item A formative study (N=8) that summarizes practices, challenges, and expectations in conducting data analysis with LLM agents based on conversation.
      \item\rev{A novel design that facilitates monitoring and steering LLM-generated data analysis script featuring interactive visualizations.} We implement a prototype system named \system and evaluate its usability (N=12).
      \item Discussions and implications on user interface design of LLM agents for data analysis tasks.
\end{itemize}

\section{Background \& Related Work}
Here, we review NLI-based data analysis tools, visualization techniques for data processing scripts, and user interface design for human-LLM interactions, which are closely related to our study.

\subsection{Demystifying NLI-based Data Analysis}
NLI-based data analysis tools interpret users' instructions in natural language and automatically perform analytic tasks.
Existing tools often assemble atomic data operations based on a clear categorization of analytical tasks~\cite{shen2022towards, chen2022crossdata}.
To support more flexible user tasks, there has been surging interest in applying LLMs to translate NL-based user intents into data-related operations or directly synthesize visualization programs (\eg~\cite{tian2024chartgpt, liu2023jarvix, liu2024ava}).

However, it remains unrealistic to expect completely correct outputs for reasons like language ambiguity and algorithmic or model accuracy~\cite{feng2023xnli, ferdowsi2023coldeco, narechania2021diy}. 
This issue becomes more pronounced when integrating LLMs into data analysis tools, given their black-box nature.
 This characteristic calls for rigorous inspection and verification strategies, as highlighted in prior research~\cite{chopra2023conversational, podo2024vievallm, gu2024wizard}. 
Example errors include wrong column selection, data mapping, data transformation, \etc
In response to the challenge, XNLI~\cite{feng2023xnli} provides a standalone interface that shows one user query to the key aspects in a finite set of the traditional NLI pipeline, \ie~attributes, tasks, and visual encodings.
With LLMs, Huang~\ea~\cite{huang2023interactive} converted the data transformation program into a flowchart using intermediate tables as nodes.
Under a spreadsheet-based interface, Liu~\ea~\cite{liu2023what} proposed \textit{grounded abstraction matching} (GAM) that explains LLM-generated code to end users in natural language.
ColDeco~\cite{ferdowsi2023coldeco} further augments GAM with two complementary views of intermediate results, highlighting how the operation changes the result.

Our work applies to analytic tasks that are more open-ended and concern complex data operations, which is under-examined~\cite{he2023text2analysis}. Most relevant to our interest in a conversational interface, Gu~\ea\cite{gu2024how} added a side panel that profiles intermediate data to facilitate retrospective examination of the synthesized code.
Kazemitabaar \ea~\cite{kazemitabaar2024improving} proposed to afford editable assumptions, execution plans, and code in LLM response for close verification and steering.
We complemented their design by proposing a transformable representation of the code, aiming to lower the abstraction level of the code and enhance user engagement during the interaction.

\subsection{Sense-making of Data Processing Code}
Simplifying data processing code can support learning ~\cite{lau2023teaching}, collaborative work~\cite{pu2021datamation}, and quality control~\cite{xiong2022visualizing, shrestha2023detangler}.
To give a comprehensive view, prior research has condensed the operations into descriptive narratives~\cite{feng2023xnli, liu2023what} or schematic diagrams~\cite{huang2023interactive, ramasamy2023visualising}.
In addition, many works focused on visualizing interim results through animation (\eg~\cite{khan2017data, pu2021datamation, guo2023urania}) or a timeline representation (\eg~~\cite{niederer2017taco, bors2019capturing, lucchesi2022smallset}).
For instance, Datamation~\cite{pu2021datamation} visually maps and links each step of the data process to the underlying dataset, providing more context for the audience.
Smallset Timeline~\cite{lucchesi2022smallset} intelligently selects samples affected by the operation and encodes the changes on a table along the timeline.

\rev{To enhance understanding of atomic data operations, many works investigated step-wise examination of the underlying data.
This can be achieved by revealing the connections and discrepancies between the input and output states.
Pandas Tutor~\cite{lau2023teaching} highlights selected rows and links their new position with arrows.
SOMNUS~\cite{xiong2022visualizing} presents 23 static glyphs for data transformation operations in table, column, and row granularity, respectively.
To bridge the mental map between data transform specifications and results, some works allow interactive inspection~\cite{kendal2011wrangler, shrestha2021unravel, shrestha2023detangler}.
For instance, Unravel~\cite{shrestha2021unravel} automatically transforms individual data operations into summary boxes with key parameters and the table size, which serves as an intermediate layer for users to modify and access runtime execution results.}

\rev{\system addresses a new problem: sense-making of data processing code produced by an LLM agent.
Compared to previous approaches that deal with complete and static scripts, the code is generated in a streaming manner, which may present challenges for users in terms of following the LLM's response during the generation process.
In addition, some tools (\eg~\cite{wang2022diff, shrestha2021unravel}) require coding proficiency while some have a rigid functionality (\eg~\cite{xiong2022visualizing, feng2023xnli}). 
However, in our scenario, end-users, including data analysts, laypeople, \etc, talk to an LLM agent for various data analysis tasks. 
We prioritize intuitive visualization designs for immediate understanding and rapid verification, keeping users engaged and undistracted during the active code generation phase.
General code debugging, however, is beyond our scope.}

\subsection{Advancing UIs for Human-LLM Interaction}
Amidst the wave of LLMs, the HCI community has been advancing user interface design to enhance control over LLMs, moving beyond a standard chatbot framework or basic API invocations.

Similar to our motivation to facilitate easier comprehension and verification of the generated content, some works seek to bridge the gulf of envisioning in human-LLM interactions~\cite{subramonyam2024bridging, tankelevitch2024metacognitive}.
For example, Graphlogue~\cite{jiang2023graphlogue} converts linear text into a diagram that encodes logical structure on the fly to assist information-seeking tasks.
Zhu-Tian~\ea~\cite{chen2024sketch} foreshadows LLM-generated code incrementally and instantly during prompt crafting.
Sensecape~\cite{suh2023sensecape} empowers users with a multilevel abstraction of existing conversation and supports information foraging and sense-making. 
We attend to an emerging scenario of conversational data analysis with LLMs, where we present novel features like on-the-fly visualization as code streams in, code scrolly-telling, and snippet navigation.

Another stream of research explores novel interaction designs with LLMs that surpass the conventional single-text prompt, where more dynamic and progressive workflows and interaction modalities are promoted.
For instance, Wu~\ea~\cite{wu2021aichains} introduced the concept of AI Chains, where users specify how the output of one step becomes the input for the next, resulting in cumulative gains per step.
Many works targeted specific application domains, including writing~\cite{chung2022talebrush}, graphics design~\cite{masson2023directgpt}, programming~\cite{angert2023spellburst}, \etc
Relevant to our interest in granular control of LLM-generated code, Low-code LLM~\cite{cai2024low} allows users to edit the tentative workflow synthesized by a planning LLM, thereby providing control over the generated code.
DynaVis~\cite{vaithilingam2024dynavis} leverages LLM to synthesize UI widgets to edit data visualizations dynamically.
Bearing a similar idea, our work supports user interactions with the intermediate visualization to drill down or refine the code in place for more intuitive and granular control with LLMs.

\section{Formative Study}
We conducted a formative study (N=8) to better understand the glitches in LLM-powered data analysis tools and inform the design considerations for contextualized support.

\subsection{Setup}
\paragraph{Recruitment \& Screening}
We posted recruitment advertisements on social media and university forums.
Candidate participants were required to complete a questionnaire about their demographic information and relevant experience.
We selected volunteers who are more experienced with data analysis and familiar with LLM-powered data analysis tools.
\paragraph{Protocol}
The study consisted of a contextual inquiry (20$\sim$40 min) and a structured interview (15 min).
First, we asked participants to show their interaction history with LLM agents in data analysis tasks.
If their original dataset is available, they will also walk the moderator through the data analysis procedure while thinking aloud.
\rev{For five participants with the original dataset at hand, we asked them to replicate one analysis session directly while thinking aloud.}
The interview ended with a list of questions regarding the overall experience.
Each participant is compensated with \$12/hour.
\paragraph{Participants}
We recruited 8 participants in total (P1--P8), with 3 females and 5 males, aged from 20 to 30.
Specifically, there are 6 postgraduate students, 1 undergraduate student (P3), and 1 data journalist (P4).
All are familiar with the data analysis mode (formally named as ``Advanced Data Analysis'' or ``Code Interpreter'') embedded in OpenAI's ChatGPT~\cite{openai2024ada} and had at least 5 sessions.

\paragraph{Analysis}
All interviews were video-recorded and transcribed into text.
Following thematic analysis~\cite{braun2012thematic}, the first author applied inductive and deductive approaches and derived initial categorized codes and themes.
The first three authors reviewed transcripts and important screenshots based on weekly meetings to agree on the final themes after iterations.

\subsection{Findings}
Here, we summarize the key findings from the interview study.

\subsubsection{Why do people turn to LLM-powered tools for data analysis?} 
Participants recognized the versatility of conversational LLM agents for data analysis as a significant advantage.
They have utilized it for a diversity of data-intensive tasks, including exploratory data analysis (4/8), data wrangling (4/8), confirmatory data analysis (2/8), data profiling (2/8), and data retrieval (1/8).
In addition, participants appreciated its flexibility in open-ended data analysis.
\iq{Compared with software with rigid functionalities, I enjoy the freedom here [in ChatGPT]. I can ask for an explanation based on the result, request recommendations for the next step, or insert irrelevant questions.} {\small (P6)}
Another strength of an LLM-powered data analysis tool is its low-code or no-code environment, where end users only need to describe the tasks and obtain a well-organized response in the form of code or report.
For instance, P4, who works in investigative data journalism~\cite{showkat2021where} and regularly cleans and organizes datasets from various sources, stated \iq{Having code generated from scratch saves days of my work}.
This feature was particularly valued by participants who were not proficient in coding (2/8). \iq{I no longer need to care about detailed operations and learn the APIs.} {\small (P2)}

\begin{table}[t]
\caption{Common issues in the code generated by OpenAI's ChatGPT for data analysis tasks.}
\label{tab:issues}
\centering
{\small
 \renewcommand{\arraystretch}{1.15}
\begin{tabular}{@{}p{2.5cm}p{5.5cm}@{}}
\toprule
\textbf{Issue Type} & \textbf{Detailed Behaviors of an LLM Agent} \\
\midrule
Incomplete workflow & Misses some important steps, \eg~not excluding empty value when computing means.\\
Non-existing symbols & Invoke a function, configure a parameter, or use a variable that is not defined.\\
Data transform failure & Fails to handle edge data value, \eg~accessing an attribute that does not exist in all data items.\\
Wrong columns & Selects the wrong column(s). \\
Unreasonable values & Sets parameter to an inappropriate value, \eg~using an overly high threshold for outliers.\\
\bottomrule
\end{tabular}
}
\Description{Table 1 lists common issues encountered by interviewees when using ChatGPT for data analysis tasks. The first column, "Issue Type," categorizes the problems, while the second column, "Detailed Behaviors of LLM Agent," provides specific examples.}
\end{table}

\subsubsection{How do people work with LLM-powered tools in data analysis?} We categorize participants' workflows into three phases: code generation, post-verification, and iterative refinement.

By default, \ADA collapses the code and communicates the progress in percentage only.
Correspondingly, participants (7/8) hardly toggled the code panel during the generation phase but distracted themselves by turning to personal matters or engaging in related side tasks like reviewing previous conversations.

Upon completion of the code generation, every participant consistently reviewed the textual response and, if available, the visualizations to grasp the analysis's implications.
Verifying the code's reliability was a common concern, with most (6/8) participants inspecting the generated script, especially when the data insights were important.
They would look into the entire data processing pipeline and specific parameters of individual operands.
P4 sometimes posed a \rev{validation} question to verify the code's correctness, such as requesting the mean value to see if it aligned with his prior knowledge.
When the generated code was inconsistent with expectations, participants (6/8) attempted to recalibrate the agent's direction through refined prompts.
P2 mentioned a special strategy: \iq{I try really hard to decompose the task into actionable items so that it won't be too challenging for \ADA.}
Notably, some participants (3/8) regenerated the response instead of starting a new conversation.
\iq{I am afraid to break the analysis flow with additional requirements on a small step.} {\small (P3)}
For open-ended tasks, after obtaining initial results, participants may further drill down through conversation (3/8) or turn to a local coding environment (2/8), depending on the trade-off between coding and prompting.
\iq{With the code, I can easily reuse it on a (computational) notebook.} {\small (P1)}

\subsubsection{What hinders human-LLM collaboration in data analysis tasks?}

Three themes emerge regarding glitches for users to participate in data analysis assisted by LLM agents actively.

$\diamond$\ \ \underline{Disrupted workflow negatively impacts user engagement.}
As code generation and execution are sometimes long-winded, it interrupts the analysis flow.
Most participants (7/8) would shift focus during the process instead of monitoring the generated code closely, for code is not as intuitive or accessible as natural language.
\iq{I feel exhausted when reading the code, so I'd rather leave it alone.} {\small (P1)} 
Without timely intervention, tiny errors in the code may propagate and invalidate the analysis result, precipitating a need to revisit and revise the work.
This leads to heightened frustration and a considerable waste of time, as finishing one exploratory data analysis task generally takes half to three minutes.
To avoid such prolonged dialogue exchanges, P3 explicitly requested the agent to ask for permission before generating and executing, explaining that \iq{(In this way,) I can at least take control over the direction}.  {\small (P5)}

$\diamond$\ \ \underline{Verifying raw code is mentally demanding.}
While LLMs may provide clear annotations to explain each step, many participants (7/8) still found verifying the generated code challenging.

On the one hand, reviewing the code snippet is inherently laborious and counter-intuitive, particularly when deciphering code from an external source, which can be mentally taxing.
After all, LLMs may not follow the coding styles the participants are comfortable with.
\iq{It [LLM] sometimes uses much-advanced syntax, so I ask it to write code like a freshman.}  {\small (P5)} 
Besides, LLMs may employ unfamiliar packages.
 \iq{I don't even know what the function parameter is about, let alone correct it.} {\small (P3)} 

On the other hand, LLMs may introduce various unexpected errors in the code that require careful inspection, as evidenced in the literature~\cite{feng2023xnli,chopra2023conversational, gu2024wizard}.
\autoref{tab:issues} lists example issues.
P6 noted LLM hallucinations: \iq{At first look, the logic was awfully smooth, yet the parameter was a synthesized constant. It's very tricky (to identify the issue).}
Some participants (3/8) were concerned about the finding's reliability but frustrated with limited approaches.
\iq{I am not sure if the conclusion is correct. I have a tight time budget, so I check the major steps and cross my fingers for no other issues.} {\small (P8)} 

$\diamond$\ \ \underline{Iterations can be extensively back-and-forth.}
To fix identified issues, users need to formulate instructions regarding what is wrong and how to correct the errors and then wait for another generation-execution-report cycle.
Unfortunately, this process can become time-consuming due to its trial-and-error nature and requires substantial effort to communicate the nuances of the desired analysis effectively.
Therefore, many participants (6/8) were reluctant to embrace the conversational workflow fully.
For minor issues like refining operational details, some participants (5/8) preferred to copy-paste the code to a local environment and make adaptations.
\iq{It is more convenient to reuse the code than telling ChatGPT specifically what to do.} {\small (P7)}
For major changes like adding a new processing step, they were more willing to communicate with the LLM agent since writing code becomes tedious. Still, after several trials, they would turn to the local environment when losing patience.

\subsection{Design Considerations}
Informed by the formative study, we draw the following design considerations (DC) to guide our conception of an alternative interaction design for LLM-powered data analysis tools.
\rev{\textbf{Our design goal is to support monitoring and steering LLM-synthesized data analysis with interactive visual scaffolding}}.

\textbf{DC1. \cut{Identify key data operations on the fly in the code generated by LLMs}\rev{Abstract code stream into key data operations for a focused verification}.}
In the context of LLM-based data analysis, a primary challenge emerges due to the often extensive and complex nature of the generated code. 
However, users usually prefer understanding the analysis process itself over the complex details of the code.
\rev{Echoing a previous study~\cite{gu2024how}, participants expressed the need to access data operations, determinant parameters, and their outcomes.}
\rev{To address this challenge, we propose to simplify the information to digest for verifying the data analysis procedure conducted by LLMs.}
\rev{By extracting the layered information concerning individual data operations from the code, such as the parametric specifications and execution results, we aim to refocus users' attention on the analysis process itself, sparing them from the overwhelming task of understanding the raw code.}

\textbf{DC2. \cut{Scaffold data analysis with intuitive visual representations faithful to the code}\rev{Scaffold data operations and execution results through straightforward visualization generated on the fly}.}
Despite the abstraction, users, particularly those with limited programming expertise, may still find it challenging to interpret the raw, syntax-heavy output produced by LLMs. 
Drawing inspiration from previous works in code visualization~\cite{myers1990taxonomies, victor2011up},
we adopt visual representations that abstract away from specific code syntax to facilitate quick comprehension of the data analysis process.
Thus, the visual representation should also expose this information, including the data state before and after each operation.
\rev{Moreover, this process should be executed on the fly along the code generation process, ensuring a seamless experience for the user aligning to their sense-making process.}
It is also critical to establish a connection between the code and its visual representation.
This will allow users to see the direct impact of their instructions on the data and to navigate the analysis workflow more effectively.

\textbf{DC3. \cut{Support granular iterations on step-wise operations} \rev{Support interrogation to the LLM and iterative code generation in the visualization.}}
An outstanding issue of LLM-powered data analysis in a conversational interface is the tediousness of articulating refinement intents and uncertainties in LLMs' follow-up responses.
To overcome this, the visual representations should simplify articulating these intents by providing mechanisms to modify the data analysis process at a granular level.
Users should be able to interact with individual steps (data operations) of the generated analysis, allowing them to make precise adjustments without the need to rewrite large portions of code or restart the conversation.
This granular control empowers users to fine-tune the analysis, accurately reflects their intentions, and streamlines the iterative refinement process.

\begin{figure}[t]
\centering
\includegraphics[width=\linewidth]{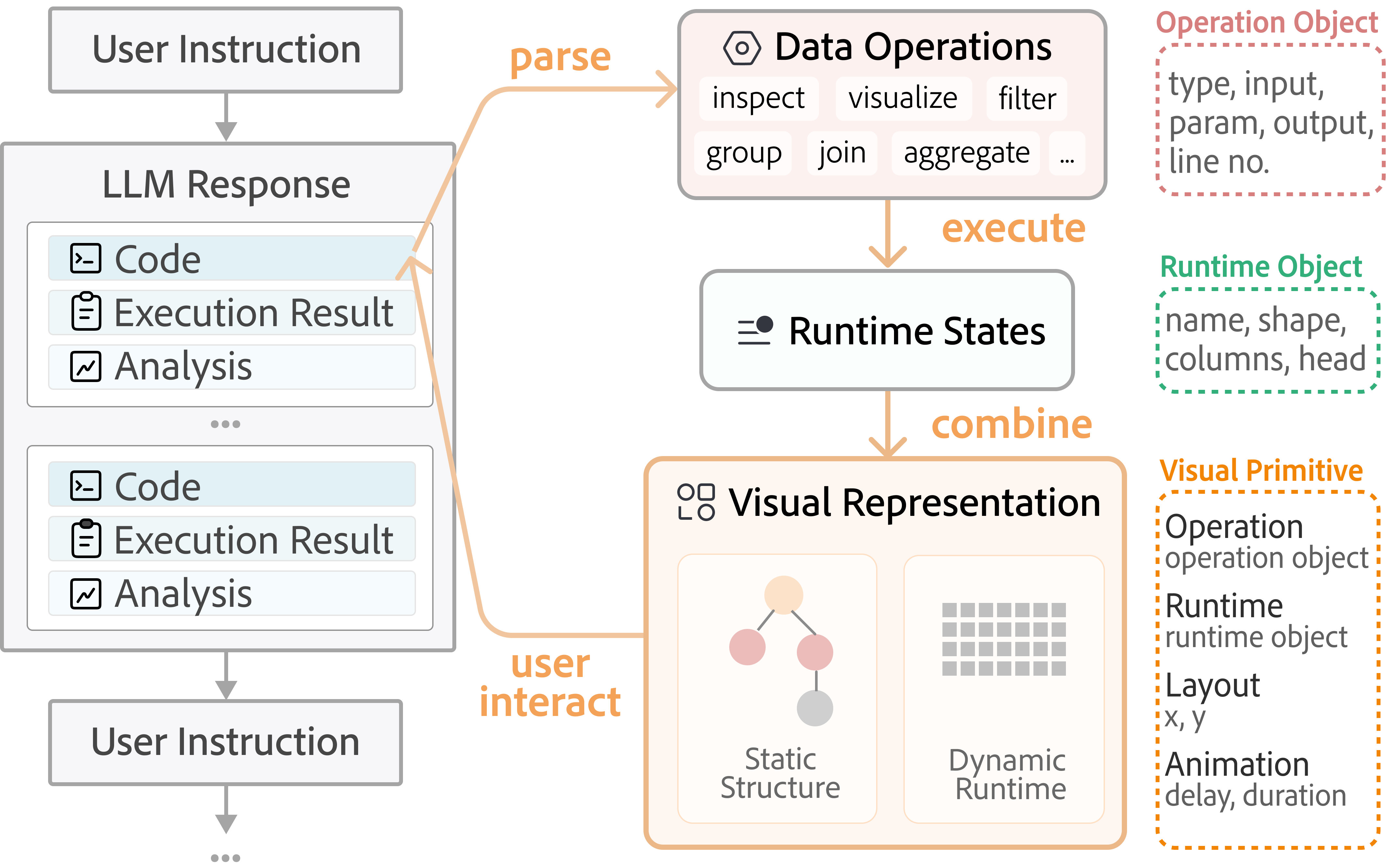}
\caption{We propose a workflow that identifies data operations within the generated code and maps them to visual, interactive primitives on the fly. 
These primitives collectively offer an overview of the data analysis process.}
\label{fig:workflow}
\Description{Figure 2 illustrates the proposed workflow. The left side shows the user-LLM interaction: users input instructions, and the LLM responds with blocks containing code, execution results, and analysis. The middle section depicts the visualization workflow: First, the code is parsed into data operations. Second, these operations are executed to derive runtime states. Third, the runtime states are combined with the static code structure to produce a visual representation that includes static and dynamic information. The right side describes the objects extracted and the focus at each stage of the visualization workflow, including the data operations, runtime states, and visual representations.}
\end{figure}

\textbf{\rev{DC4. Embed visualization seamlessly into the conversational user interface (CUI).}}
\rev{As conversational data analysis normally takes place in a CUI~\cite{gu2024how, chopra2023conversational}, we tailor the design to common design patterns of web CUIs in a non-intrusive manner.
For instance, the visualization should be stably revealed during the progressive generation, following the same vertical order as the code.
It should offer a lightweight complementary view of the code section in the LLM's response (see \autoref{fig:workflow}) and afford a level of visual guidance for the code dependency between conversational threads.
}

\begin{figure*}[th]
\centering
\includegraphics[width=\linewidth]{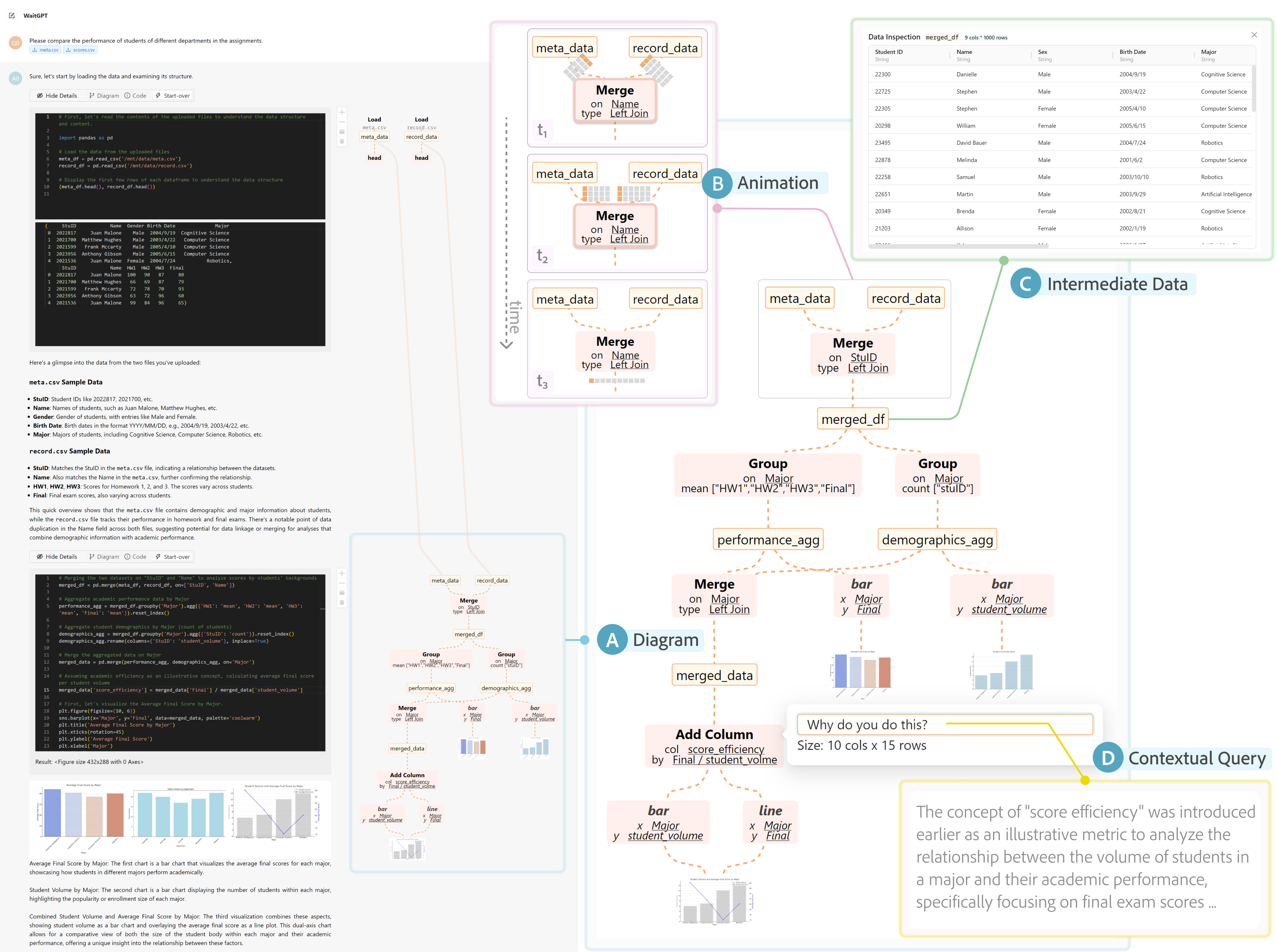}
\caption{\rev{A screenshot of the WaitGPT user interface. (A) An enlarged view of the flow diagram representing the code. (B) An illustration of the ``table glyphs'' that flow along the edge showing table dependency and changes during code generation. (C) Inspecting intermediate data by toggling the interactive table panel. (D) Interrogating LLM based on an operation.}}
\label{fig:interface}
\Description{Figure 3 shows a screenshot of the user interface. The left side presents an overview of a conversation, while the code visualization is enlarged on the right side. There are three subfigures. The top subfigure showcases animated glyphs during code generation. The middle-right subfigure illustrates an interactive table view. The bottom-right subfigure demonstrates contextual inquiry based on a specific operation.}
\end{figure*}

\section{\system: Usage Scenario}
\label{sec:walkthrough}
Informed by the formative study and design considerations, 
we propose dynamically visualizing the code generation process to help users steer 
a conversational LLM agent during the data analysis process.
This is achieved through a workflow that identifies data operations within the generated code and maps them to visual primitives on the fly (see \autoref{fig:workflow}). 
These visual primitives not only illustrate the static aspects of data operations but also display the runtime states of the underlying data (\ie~tables) both before and after these operations. 
Moreover, they provide users with rich interaction possibilities, allowing them to refine the data operations without regenerating the code entirely.

We instantiate this idea with a prototype system, \system, 
which enables users to proactively guide the data analysis process with an LLM agent, 
making interventions akin to saying, {\it ``Wait, GPT, there is something wrong...''}
This section walks through \system using a hypothetical use case, 
demonstrating its capacity to transform the user's interaction with LLMs in data analysis tasks.

\paragraph{\textbf{Usage Scenario}}
\user, a college lecturer, would like to review her students' performance across assignments to inform future teaching strategies. 
She opened \system, an LLM-powered conversational tool for data analysis that she was familiar with.

\system's interface resembles a chat box, allowing users to upload spreadsheets and inquire about the data in natural language (\autoref{fig:interface}).
Upon uploading two spreadsheets --- one detailing student profiles and the other their individual assignment scores --- \user asks \system to compare the performance of students with different backgrounds. 
In response, \system outlines a plan to meet her requirements, then crafts a code snippet to conduct analysis.
An external executor executes this code snippet to yield results. 

Unlike similar tools, \system visualizes the data analysis process instead of just presenting raw code and textual execution results (\autoref{fig:interface} A).
It dynamically extracts data operations and presents them as nodes within a diagram illustrating the data flow. 
For instance, a ``\texttt{join}'' operation node would display as ``merge''. And the node shows the tables being joined, the type of join (\eg~left join, cross join, \etc), and the indexing column used for the join.
These blocks are linked based on dependencies and posited from left to right to reflect the procedural order.
Notably, \system breaks down the analysis script into executable blocks that are executed immediately instead of executing until the entire code snippet is ready.
This allows for a progressive understanding and debugging process, enabling users to see the effects of each operation in real time. 
The tool also visualizes the runtime state of data tables (\eg~the number of data entries/columns, selected columns) as part of the diagram.
Specifically, the runtime state of each table is visualized as glyphs, which move along the linked edges between operation objects.

Through the visual representation, \user quickly spots a flaw in the diagram—the row number reduces (\autoref{fig:interface} B).
Rather than requiring rewriting the original query and regenerating the entire data analysis code,
\system enables users to refine specific operations directly within the visualizations. 
Users can directly update its parameters, inquire about details, and indicate refinement intents through natural language.
Thus, \user adjusts the join parameters to student IDs, and then clicks on the re-run button to execute the updated code.
While the analysis goes on, \user inspects the table.
She requests the LLM to clean the data. The diagram updates, reflecting the corrected scores after the agent integrates a data validation operation.
Now \user is ready to analyze the reliable data, her teaching plans are secure on a foundation of accuracy.

\section{\system: System Design}

The design of \system consists of three major components: abstracting the code to data operation chains, visualizing these chains, and providing interactions to steer the analysis process.

\subsection{Abstracting Code to Operation Chains}

\rev{Based on the interview, we identified three types of information indispensable for code comprehension: table variables, data operations, and execution results.
In addition, different data operations encapsulate dedicated semantics and independent parameters.
Therefore, we opt to abstract a data analysis process into a graph structure, chaining its nodes with an input-output relationship as follows (\textbf{DC1}).}
The input of each data operation is table(s), 
whereas the output can be the updated table, new table(s), 
other derived values/visualizations, or none.

\begin{itemize}[label=$\ast$]
    \item \textit{Table node}: A table node corresponds to a variable for an underlying table in the code, such as a \texttt{dataframe} in the Pandas package.
    It can be either loaded from a data file or dynamically generated during code execution as an interim variable.
    
    \item \textit{Operation node}: An operation node ties to an atomic data operation. It surfaces the detailed parameters of an operation object, \eg, \op{Select}, \op{Filter}, and \op{Sort}.
    
    \item \textit{Result node}: A result node is associated with an execution result, such as printed values or data visualization.
\end{itemize}

Additionally, the relationship between these nodes can be one of the following:

\begin{itemize}[label=$\ast$]
    \item \textit{Input}: From table node(s) to an operation node. It means the data operation is based on the input table(s).
    
    \item \textit{Assignment}: From an operation node to a new table node. It means a new table-typed variable is yielded from the operation.
    
    \item \textit{Result generation}: From an operation node to a result node. It means the operation outputs some visible results.
    
    \item \textit{Operation chain}: From an operation node to an operation node. It means a table undergoes the two operations sequentially.
\end{itemize}

\paragraph{Extracting the Nodes through Static Analysis.} 
To extract these nodes and relationships,
we perform static analysis on the abstract syntax tree (AST) of the generated code, 
where we apply heuristics informed by patterns of data analysis scripts and functional interface design of relevant packages.
\system currently can parse atomic operations including \op{Load Data}, \op{Inspect},  \op{Select}, \op{Filter}, \op{Sort}, \op{Transform},   \op{Group}, \op{Aggregate}, \op{Merge}, \op{Add} \\ \op{Column}, and \op{Visualize}, based on the \texttt{Pandas}, \texttt{Matplotlib}, and \texttt{Seaborn} packages, which are the default choices of ChatGPT and widely adopted~\cite{chen2024viseval}.
For instance, \lstinline{merge_df = df[["attr_1", "attr_2"]].sort()} will be converted into two operation objects: \op{Select} and \op{Sort}.
To bind the table targets to the operations, we maintain a global variable of existing table variables.
This is because a table variable can only be created by being loaded from external sources (files, database, \etc) or generated as the output of prior operations.

\subsection{Visualizing Data Operation Chains}
Our goal is to transform the LLM-generated code into easily interpretable visualizations, 
facilitating user inspection of the data analysis process (\textbf{DC2}). 
To this end, we have developed a suite of visual primitives,
which present the details of each operation and their internal runtime states.
These primitives are chained together, collectively offering an overview of the data analysis process.

\paragraph{Visual primitives for the static code}
We utilize a diagram to represent the graph-based data processing procedure for individual code snippets.
The table node, operation node, and result node are visualized as blocks, color-encoded in yellow, pink, and white.
\rev{A node-style visualization is chosen for its familiarity to general users (\textbf{DC2}) and flexibility in displaying layered information, expanding with the code stream, and implying the operation order (\textbf{DC4}).}
As LLMs sometimes synthesize long variable names for clarification, we considered a rectangular block beneficial for encapsulating this information.
For simplicity, a table node only shows the corresponding variable name, and a result node shows a thumbnail.
For an operation node, we use a bold font style to prioritize the communication of its type (\eg~filter, group, \etc).
And we visually differentiate its parameters' names and values through typography.

An operation chain spans from top to bottom, following its procedure.
For a table node, there can be multiple associated operation chains.
These chains are aligned from left to right with respect to the execution order.
A code snippet depends on preexisting code as the runtime environment is shared throughout a conversation.
Therefore, a table node may trace back to previous snippets.
To reflect such a relationship, a copy is made in such a situation, which is linked to its previous occurrence with a cross-conversation curve.

\paragraph{Visual primitives for the runtime states}
The diagram is further enriched by visual glyphs that encode the runtime status of table variables.
A table glyph takes a common visual representation for tables---a 2D matrix.
The number of matrix columns is the same as the column number of the table.
The number of matrix rows per column is proportional to the number of table rows \rev{to roughly indicate changes in data size and scale to different data sizes.
To access precise information about the runtime states, one may interact with the associated operation node for details.}
Through chained operations, the size of a table can be updated.

\subsection{Steering the Data Analysis of LLMs}

The diagram goes beyond merely a visual representation of the data analysis process.
It also acts as an interactive scaffold for users to steer data analysis code generated by LLMs, 
enabling real-time inspection, retrospective examination, and granular refinement (\textbf{DC3}).
This section introduces interactions supported in \system.

\subsubsection{Real-Time Inspection on the underlying code}
\begin{figure}[t]
\centering
\includegraphics[width=\linewidth]{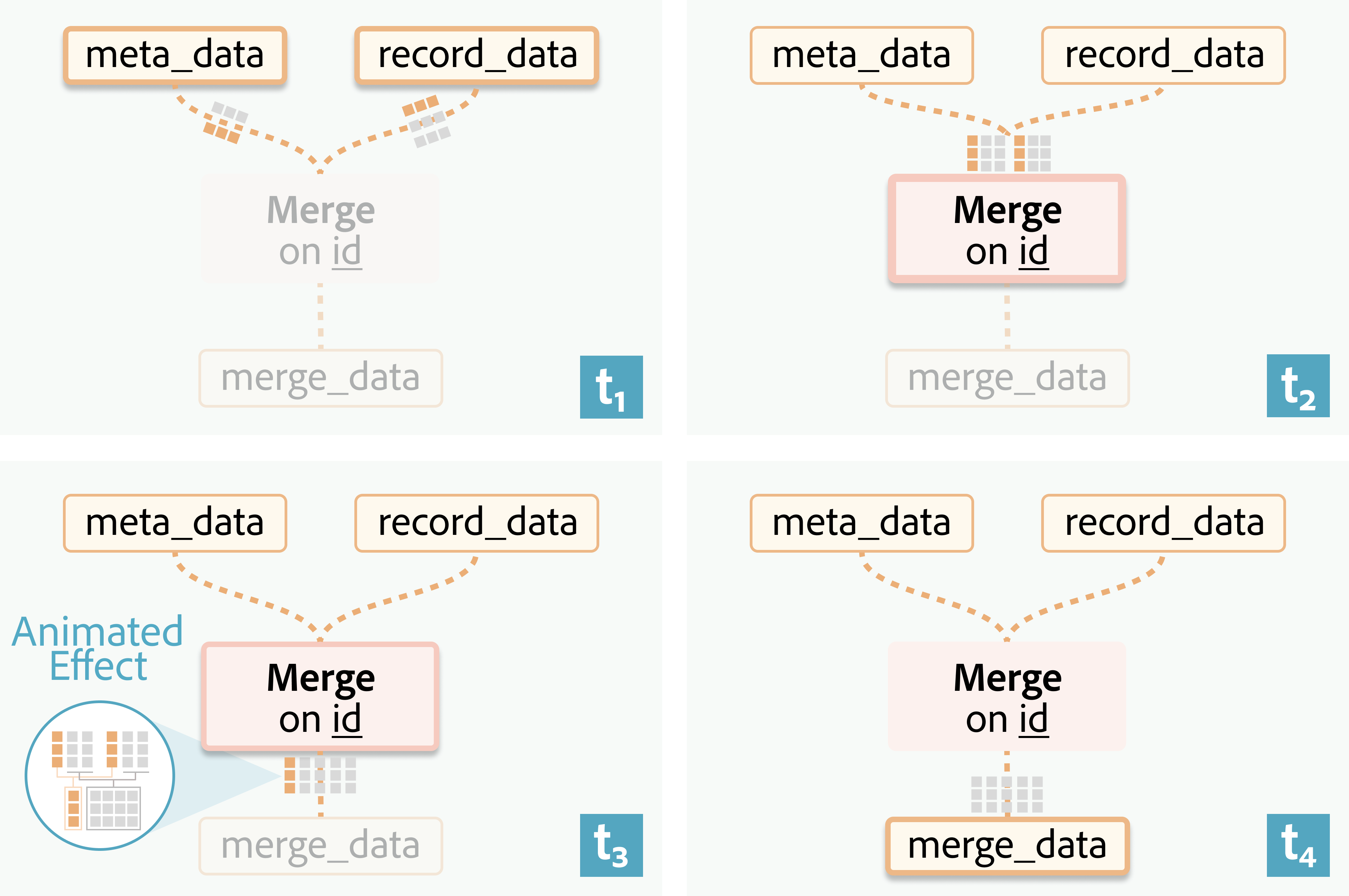}
\caption{An illustration of how the diagram grows with animated table glyphs during the code generation process.}
\label{fig:animation}
\Description{Figure 4 uses four subfigures to demonstrate animated keyframes of a flow diagram during code generation. The first subfigure shows that two datasets are loaded. The second subfigure shows the two datasets being merged based on a mutual column named "id". The third subfigure shows the merged dataset, highlighting the animated effect. The last subfigure shows that a new data frame named "merge_data" is created.}
\end{figure}

During code generation, only the diagram is shown to reduce the cognitive load of end users.
However, they may still toggle on the code panel and juxtapose the diagram side-by-side.
When a data operation is being activated, \ie~the external executor has just run the code, it will be added to the diagram, potentially introducing a new table node or a result node.
Meanwhile, relevant table glyphs also appear and gradually flow from the previous node to the current node. 
\autoref{fig:animation} showcases an example of the dynamic process.

\subsubsection{Retrospective Investigation on the analysis process}
After the code and diagram are completely generated, users may perform a retrospective examination to verify the procedure and investigate potential issues.
\rev{To evaluate the analysis flow, users may replay the animation showing diagram expansion or utilize scrolly-telling, where they can take control over the animation progress using scroll-based interactions (\textbf{DC4})}.
If the code panel is toggled on, the corresponding line(s) of code will be highlighted for activated nodes upon re-play or the user's mouse hover events (see \autoref{fig:interaction} A).
This feature bridges the visual representation and the textual code, visual navigation and troubleshooting.
Essentially, nodes in a diagram are visually displayed in the simplest way to support fast comprehension.
To access details about the underlying data tables in the runtime context, users may click on a node of interest and review an additional panel (see \autoref{fig:interaction} B). 
The thumbnail of a visualization result node is expandable (see \autoref{fig:interaction} E).

\subsubsection{Granular Refinement}
\begin{figure}[t]
\centering
\includegraphics[width=\linewidth]{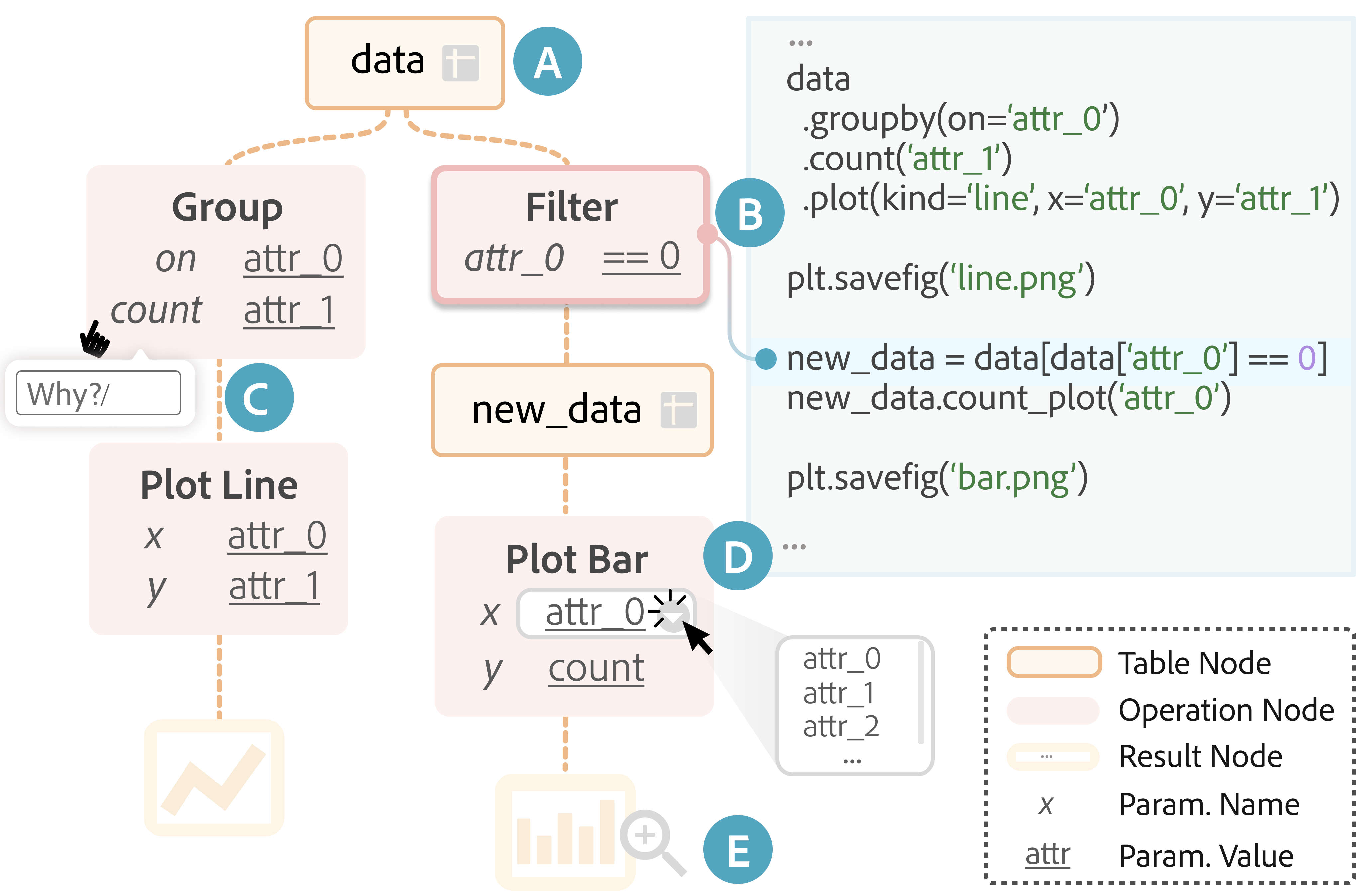}
\caption{The visualization offers multiple interactions for inspecting and refining the underlying data analysis. Users can: (A) toggle a table node to view the underlying data; (B) hover over a node to highlight its corresponding code; (C) modify a data operation using natural language; (D) directly manipulate the parameters of a node; and (E) view the resulting visualizations from the analysis. 
}
\Description{Figure 5 contains three components: the flow diagram on the left, the associated code on the top right, and the legend on the bottom right. The analysis diagram on the left shows code execution linked to visual elements. "A" marks the initial table node, splitting into two paths: Path one goes to a group operation node, then to a plot line operation node, and finally to a line chart. A hand icon at the group operation node prompts a "Why?" inquiry labeled "C". Path two leads to a "Filter" operation node, a "new_data" table node, and then to a "Plot Bar" operation node, resulting in a bar chart. A line from the "Filter" node to a specific line in the code is labeled "B". In the "Plot Bar" node, a dropdown menu with a cursor clicking on it is shown, labeled "D". A magnifier icon next to the bar chart indicates an interactive feature labeled "E".
}
\label{fig:interaction}
\end{figure}

The diagram offers new interaction modes for granular refinement through direct manipulation and contextual interrogation.
\rev{Instead of regenerating the entire analysis, which may involve multiple code snippets, users can steer the data analysis at a finer granularity within the visualization (\textbf{DC3}).}
Users may directly manipulate the operation objects based on their visual representation and update the underlying code (see \autoref{fig:interaction} D).
The fields of parameters in operation nodes are editable input forms, allowing fine-grain updates.

Similar to the concept of interrogative debugging~\cite{ko2004designing},
users can select specific operation nodes within the diagram and then request explanations or suggest revisions to the LLM by focusing on a particular node, which offers a targeted context for verification and refinement (\autoref{fig:interaction} C).
This provides an alternative mode to the common practice of selecting code or table cells and posting queries to LLMs~\cite{nam2024using}.
Inspired by the regeneration practice of participants in the formative study, the query is independent of the main conversation and, thus, will not affect the memory of LLM agents.
The LLM's suggestion of code update will directly apply to the code panel, and the previous version will be commented out for comparison.
When satisfied with the refinement, users can re-run the code snippet to attain updated analysis from LLM agents based on the new execution results.

\section{Implementation}
\system is a web-based app implemented in the React~\cite{react} framework based on TypeScript.
\rev{We apply the Monaco Editor~\cite{monaco} to display the code with standard syntax highlighting.}
We adopt the OpenAI's API, with the  \texttt{gpt-4-0125-preview} model.
To manage user-uploaded files, parse LLM-synthesized code into an abstract syntax tree, and obtain its execution result, we also host a back-end server implemented in Python with Flask~\cite{flask}. 
The LLM prompts applied in \system generally follow the guidance of OpenAI with little engineering effort.
Our implementation integrates three key mechanisms as follows.

\paragraph{Session Management}
In addition to the conversation history for each session, \system maintains other contexts to support diagram generation on the fly and granular refinement.
The associated contexts include a sandbox environment for file storage and code execution, a global record of table variables, and specifications of the diagram for each data analysis code snippet.
In addition to the parsed parameters, the runtime status of target tables, and rendering configurations, the specification of a data operation node in a diagram also records conversation logs with the LLMs based on the code to support iterative refinement.

When a user sends a query, the LLM will respond with textual contents or a function call to the pre-declared Python executable.
For code-based response, \system first decides whether it is about data analysis and then activates the automatic parser.
The runtime context for each code snippet is cloned from the main process and cached for potential rework, thus enabling flexible user interruptions and refinement at any point.
We enhance user navigation by prompting LLM to summarize the main task and build a minimap for existing data analysis snippets.

\paragraph{Sandbox Execution}
Before running the code in a sandbox environment, \system refactors the method chain into separate standalone statements.
Therefore, based on the identified targets (\ie~table variables) of data operations, the static parser inserts printing statements based on templates to retrieve the intermediate status of the table, including the number of rows, the number of columns, and column names.
The table status is then bonded to the corresponding data operation object.
As a note, we opt to insert code to the LLM-generated script in a post hoc manner to reduce dependency on specific versions.
An alternative approach is to inject logging facilities into the standard libraries~\cite{pu2021datamation, shrestha2021unravel}.

\paragraph{Rendering} 
The rendering of the flow diagram comprises two steps.
Once the static analyzer extracts new data operation objects, they will be added to the diagram using a graph layout algorithm and maintain inactivated status.
When the runtime information is bound to the operation object, its animated effect is pushed to a queue to play sequentially, where the corresponding node will be activated and the table glyph will flow from the prior node to the current node.

\section{User Evaluation}
We evaluate \system through an in-lab user study with 12 participants of various backgrounds and data analysis expertise.
Specifically, we are interested in the following research questions.

\begin{itemize}

    \item How effectively does \system facilitate intermediate verification during the generation process of LLM agents?
    \item How effectively does \system support retrospective verification after data analysis tasks are completed?
    \item To what extent does \system support the granular refinement of generated code snippets?
    \item How do users perceive the usefulness of \system in their daily data analysis tasks?
\end{itemize}

\begin{table*}[t]

\caption{The success rate (\%) and average duration (seconds) in \system and \baseline for Task A \& Task B (N=6/condition). The failure column describes the mistake made by LLMs in the task. \#Line: No. lines in the code snippet; \#Char: No. characters. \#Df: No. table nodes in the data operation chains, \#Op: No. operation nodes, \#Res: No. result nodes. ``(Value)'': standard deviance. }
\label{tab:performance}
\centering

  \setlength{\aboverulesep}{0.5pt}
\setlength{\belowrulesep}{0.5pt}
\begin{tabular}{clrrrrr|r|r|r|r}
\toprule
\multirow{2}{*}{\textbf{Task}} & \multirow{2}{*}{\textbf{Failure}} & \multirow{2}{*}{\textbf{\#Line}} & \multicolumn{1}{l}{\multirow{2}{*}{\textbf{\#Char}}} & \multirow{2}{*}{\textbf{\#Df}} & \multirow{2}{*}{\textbf{\#Op}} & \multirow{2}{*}{\textbf{\#Res}} & \multicolumn{2}{c|}{\textbf{Success (\%)}}                           & \multicolumn{2}{c}{\textbf{Average Duration (s)}}                 \\ \cline{8-11}
                               &                                   &                                  & \multicolumn{1}{l}{}                                 &                                       &                                       &                           & \multicolumn{1}{c|}{WaitGPT} & \multicolumn{1}{c|}{Baseline} & \multicolumn{1}{c|}{WaitGPT} & \multicolumn{1}{c}{Baseline} \\
                               \midrule
\textbf{A1}                    & Sort on string                    & 14                               & 474                                                  & 2                                     & 5                                     & 0                         & 83 (0.41)       & 33 (0.52)      & 65.83 (45.32)        & 136.67 (88.69)      \\
\textbf{A2}                    & Miss a group condition            & 5                                & 233                                                  & 2                                     & 3                                     & 0                         & 50 (0.55)       & 50 (0.55)      & 88.33 (40.21)        & 102.50 (68.68)      \\
\textbf{A3}                    & NA                                & 47                               & 1,836                                                & 5                                     & 10                                    & 1                         & 100 (0.00)      & 100 (0.00)     & 154.00 (103.00)      & 151.67 (143.69)     \\
\textbf{A4}                    & Miss a filter condition           & 10                               & 509                                                  & 3                                     & 4                                     & 0                         & 67 (0.52)       & 83 (0.41)      & 86.67 (18.62)        & 92.50 (34.31)       \\
\textbf{A5}                    & Miss dropping duplicates          & 24                               & 780                                                  & 1                                     & 5                                     & 1                         & 50 (0.06)       & 50 (0.55)      & 92.50 (58.37)        & 87.50 (27.34)       \\
\textbf{A6}                    & NA                                & 21                               & 817                                                  & 1                                     & 3                                     & 1                         & 100 (0.10)      & 100 (0.00)     & 144.17 (69.02)       & 95.83 (22.45)       \\
\textbf{B1}                    & NA                                & 29                               & 1,167                                                & 4                                     & 7                                     & 1                         & 100 (0.00)      & 83 (0.41)      & 141.67 (49.97)       & 160.00 (82.16)      \\
\textbf{B2}                    & Miss dropping duplicates          & 25                               & 1,287                                                & 5                                     & 6                                     & 1                         & 67 (0.52)       & 50 (0.55)      & 242.50 (167.74)      & 221.67 (159.80)     \\
\textbf{B3}                    & NA                                & 25                               & 1,262                                                & 4                                     & 6                                     & 1                         & 100 (0.00)      & 100 (0.00)     & 185.00 (113.31)      & 176.67 (64.94)     \\
\textbf{B4}                    & Wrong aggregation logic           & 10                               & 654                                                  & 4                                     & 6                                     & 0                         & 83 (0.41)       & 83 (0.41)      & 212.50 (145.87)      & 138.50 (40.71)    
  
   \\
\bottomrule
\end{tabular}
\Description{Table 2 is structured into eleven rows (one header row and ten tasks) and nine columns, providing a side-by-side comparison between WaitGPT and Baseline systems across ten tasks (A1-A6 and B1-B4). The header row outlines the column titles: "Task" for task identifiers, "Failure" detailing mistakes made by LLMs, "#Line" for the number of code lines, "#Char" for the character count, "#Df" for the number of table nodes, "#Op" for operation nodes, and "#Res" for result nodes. The "Success (\%)" and "Average Duration (s)" are split into two columns each to compare the performance of WaitGPT and Baseline. Standard deviations for success rates and average durations are provided in parentheses.
}
\end{table*}

\subsection{Participants}
We recruited 12 participants (10 males, 2 females; ages 23---30, M = 26.33, SD = 2.15) through social media and word-of-mouth.
They were postgraduate students with diverse backgrounds in databases, machine learning, visual analytics, industrial engineering, computational sociology, and HCI.
According to their self-rating based on a 5-point Likert scale (1: lowest extent, 5: greatest extent), participants were generally adept at data analysis (M = 3.67, SD = 1.37) and familiar with the Pandas syntax used in \system (M = 3.5, SD = 1.38).
They were experienced with LLM-powered chatbots (M = 3.75, SD = 1.06).
Specifically, 5/12 participants leveraged ChatGPT to analyze more than 20 datasets, whereas 4/12 analyzed less than 5 datasets on ChatGPT.

\subsection{Protocol}
\paragraph{Tasks}
There are three tasks in total.
Task A is based on the \textit{Employee} dataset\footnote{\url{https://www.kaggle.com/datasets/soorajgupta7/corporate-compensation-insights}} with six analysis tasks (A1--A6).
Task B is based on the \textit{Flight} dataset\footnote{\url{https://www.kaggle.com/datasets/shubhambathwal/flight-price-prediction}} with four tasks (B1--B4).
For Tasks A \& B, the participants are required to address individual questions by interacting with LLMs and decide if the LLM-generated code is error-free.
To cover representative cases, we included both confirmation and exploratory tasks on two tabular datasets and replicated 4 known errors made by LLMs~\cite{gu2024how}.
In addition, we prepared dedicated prompts for the participants to ensure that the first LLM-generated content was identical in each task.
These prompts are grounded in the ARCADE~\cite{yin2023natural} and Text2Analysis~\cite{he2023text2analysis} datasets.
Each data analysis task is independent of the other, including common data insight types~\cite{ding2019quickinsights}, \eg~rank, distribution, outlier, \etc
\rev{Task C is based on the synthesized dataset used in the usage scenario (see \autoref{sec:walkthrough}}), where participants were asked to explore the dataset freely.
We also offer a list of self-curated queries for their reference.

\paragraph{Baseline and Apparatus}
We removed the extended view of the diagram as the baseline system, namely \baseline.
\baseline retains essential functionalities of ChatGPT that the participants are familiar with.
The code snippet offers by-line textual comments explaining each step for user verification \rev{and has standard syntax highlighting for Python.}
Meanwhile, \baseline shares the same visual appearance as \system.
This ensures that any differences in user interaction can be attributed to the diagram's presence or absence rather than other factors like aesthetics or layout.
Participants joined the study in person and finished their tasks on standardized desktop devices to eliminate hardware variability as a confounding factor.

\paragraph{Procedure}
We opted for a counterbalanced within-subjects design to compare \system and \baseline.
There are two groups (I, II) that participants were randomly assigned to.
In Group I, participants finish A1-3 \& B1-2 in \baseline, and A4-6 \& B3-4 in \system.
Conversely, in Group II, participants finish A1-3 \& B1-2 in \system, and A4-6 \& B3-4 in \baseline.
This approach allowed each participant to experience both conditions while performing a balanced set of tasks across the two systems.

The user study begins with a presentation of the visualization and interaction design, where participants can ask for details (5 min).
Then, the participant should work on Task A1-6 (15-30 min), Task B1-4 (10-20 min), and Task C (5-15 min) sequentially.
The study ends with a semi-structured interview (10-15 min) and a questionnaire (5 min).
A facilitator conducted one-on-one sessions with each participant, closely observing and taking notes of participant behaviors.
The post-study interview was audio-recorded for later analysis.
Participants were compensated with \$12 per hour.

\subsection{Measures}
We adopted the NASA-TLX~\cite{hart1988development} questionnaire to measure the perceived cognitive load in steering LLM-synthesized data analysis.
We developed a questionnaire based on a 7-point Likert scale to evaluate the usefulness of \system.
For each pre-recorded query, the facilitator records (1) the time cost that the participant discerns issues in the result since response generation, (2) the time cost that the participant makes a judgment on the correctness, (3) whether the data has been examined, and (4) whether the code panel is expanded when viewing diagrams only.

\subsection{Results}
To compare \baseline and \system, we analyze task correctness for Task A \& B and the subjective ratings of the participants.
We further report insights from the interview, 

\subsubsection{Task Correctness}

\autoref{tab:performance} lists detailed configurations and participant performance in Task A1-6 and B1-4.
In general, the success rates in the \system condition are no less than the \baseline condition, except for A4.
A4 asked for 10 employees with the highest salary currently, whereas LLM did not filter out those on leave.
Many participants did not notice this problem in the response.
As for the duration, the two conditions had similar time costs ($\le$ 10s) for Task A3-5 and B3.
And \system took less time in Task A1-2 and Task B.
However, multiple factors are attributed to the total duration, as seen in the relatively large standard deviation values.
For instance, we did not consider expertise in data analysis when assigning participants to different groups.
When the participant chose to inspect the code after viewing the diagram, there was an additional time cost to browse the code.

\begin{figure*}[t]
\centering
\includegraphics[width=\linewidth]{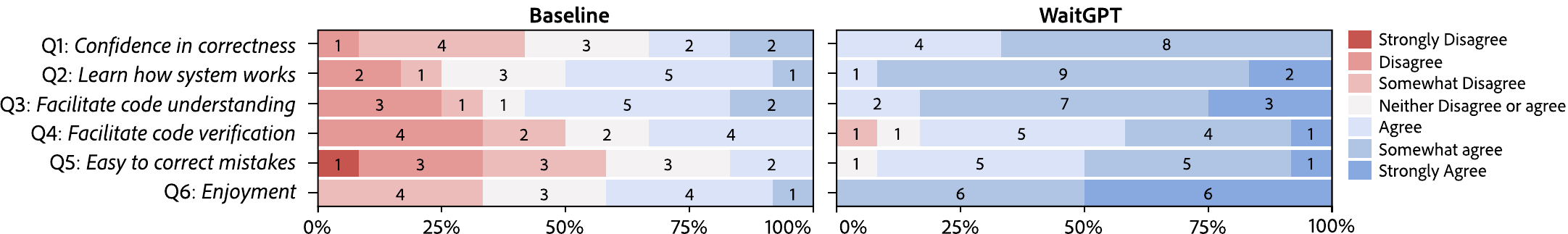}
\caption{User ratings on the baseline (code-only interface) and WaitGPT.}
\label{fig:rating}
\Description{Figure 6 shows a horizontal stacked bar chart comparing user responses for the Baseline and WaitGPT systems against six questions. The chart sequentially presents the questions on the left, the corresponding response bars for Baseline and then WaitGPT, and concludes with a legend on the right. The legend interprets the color gradient from dark red, indicating "Strongly Disagree", to dark blue for "Strongly Agree." Baseline"s bars are mostly red, suggesting neutral to negative responses. Meanwhile, WaitGPT"s bars are predominantly blue, showing a tendency towards agreement on usability.}
\end{figure*}

\subsubsection{Subjective Ratings}
As the questionnaires are based on an ordinal Likert scale and the sample size is relatively small, we performed the Wilcoxon signed-rank test to compare the subjective ratings between \baseline and \system.

 $\diamond$ \ul{On the cognitive load}.
In the NASA-TLX questionnaire, \system demonstrates lower cognitive demand to the participants.
According to the statistical tests, there are highly significant differences (p<.001) in the mental and physical demand, performance, and affective states between the two conditions.
The difference in the effort to accomplish self-performance level (p=.010) and the temporal demand (p=.050) is also significant.

 $\diamond$ \ul{On the usefulness}. \autoref{fig:rating} compares the distribution of the user ratings on  \baseline and \system based on our self-developed questionnaire.
For each question, \system attains a higher median rating than \baseline at a confidence level of 99.5\%, demonstrating its usefulness in demystifying the analysis (Q1-3), verifying or correcting the code (Q4-5), and engaging end-users (Q6).
Notably, while participants varied in task performance, 10/12 people reported increased confidence in the correctness of the analysis result (Q1).
Besides, based on a 7-point Likert scale (1: strongly disagree, 7: strongly agree), the participants considered it easy to comprehend the visualization design (Med=6.5, M=6.65, SD=.87) and interact with the diagram (Med=6.0, M=6.33, SD=.65).

\subsubsection{General impressions}  The participants were generally positive about \system and affirmed its support in monitoring and steering LLM-generated analysis. 

$\diamond$ \ul{Difference in the UX between conditions.}
Despite in-line explanations and meaningful variable names in the LLM-generated code, the participants found it mentally taxing to follow the source code and unguided in verification.
The reasons include memory demand for excessively long content (8/12), limited runtime contexts (3/12), and unfamiliar coding styles (2/12).
In comparison, participants (12/12) resonated with the ease of understanding and verifying the code in \system with a higher level abstraction. 
The diagram ``{\it strips off unimportant details}'' {\small (P5)} and offers an overview of the code.
\iq{It [the diagram] has a clean structure and can serve as a navigation for the code.} {\small (P11)}
This also kept participants engaged during the code generation.
\iq{I felt stressed viewing the code stream, but it's a pleasure to watch the diagram grow.} {\small (P4)}
The benefits of a visual summary were more apparent when the underlying code was long, as the diagram fit in the screen without the need to scroll vertically or horizontally (3/12).
Lastly, many participants (8/12) were positive about the node-based interaction instead of sending a new chat.
\iq{There's a chance that a new chat introduces new errors, so I prefer to change the code directly.} {\small (P9)}

$\diamond$ \ul{Perceived usefulness of the visualization.}
The current visual design was well-received by the participants (12/12).
We categorize the perceived usefulness of the extended visualization and associated interactions into three dimensions.

First, the diagram offers an abstract layer to focus on high-level logic and task decomposition.
As observed by P12, \iq{GPT outputs pretty code with mostly correct functional calls. This makes me lose caution for logical errors.}
P3 claimed that the visualization facilitated LLM alignment---\iq{I have a rough idea of how to process the data, and the diagram makes it easy to compare with my mind map.}

Second, the visualization surfaces information at different layers, including the detailed parameters for data operations, profiles of the data table, and navigation back to the source code.
For instance, the high accuracy rate for Task A1 was due to the convenience of inspecting data tables.
\iq{It's great to access the table right away. It's [the diagram] like an information hub.} {\small (P10)}
P3 appreciated the typography applied in the operation nodes, as \iq{it separates the variable names, operations, parameter names, and parameters}. 
P7 noted that the table glyphs suggested the semantics of unfamiliar functions through the input-output trace.

Third, the node-based interactions offer a granular approach to interrogating or modifying the code.
\iq{I prefer talking to nodes in the diagram because the context is preserved, so I don't need to type much. It's nice to have something to point to make things clearer.} {\small (P9)}

Some participants (2/11) felt more comfortable manipulating the nodes than overwriting the code.
\iq{Here [in the diagram], I don't need to care much about syntax but doing minimum updates.} {\small (P5)}
In addition, the context of a node-based interaction is constrained to the corresponding code section parallel to the entire conversation.
\iq{I am happy to maintain a clean conversation thread.} {\small (P11)}

\subsubsection{Glitches in using \system} Despite the benefits mentioned, users encountered several glitches while using the prototype.

$\diamond$ \ul{Diverse needs for level of details.} Participants had divergent perspectives on the current design of \system.
For instance, P10 expressed the hope of showing relevant annotations directly on the operation nodes.
For the table glyphs, a few participants (2/12) competent in data analysis criticized them as trivial.
\iq{I'd prefer a small annotation showing the table dimensions.} {\small (P9)}
However, some participants (3/12) embraced the design and commented that its animation double encoded the program procedure, in addition to the implicit node layout from left to right---\iq{When the code has complex dependencies, I can follow the operations step by step with the table glyphs.} {\small (P2)}
To accommodate diverse needs, a customizable interface is anticipated for flexible user configuration.

$\diamond$ \ul{Concerns in the reliability \& expressiveness.}
Participants with a computer science background (8/12) were generally interested in how the code was transformed into the diagram and expressed concerns about algorithmic failures (1/12) or potential information loss (2/12).
Like what P12 asked: \iq{What if it [LLM] made errors in parameters not presented in the diagram?} 
P6 recalled that he sometimes copied his code and prompted LLMs to use customized lambda functions for data transformation.
However, in the current implementation, \system will only tag this as a ``lambda function'' without presenting more details due to the limit of current heuristics.
As there are limited datasets on LLM-synthesized data analysis code at the moment, it remains challenging to systematically evaluate the coverage of our heuristics.
To mitigate these concerns, future improvements may incorporate automatic verification of the parsing results and generative AI to surpass expressiveness limits.

\subsubsection{Opportunities for Applications} The participants shared several creative ideas for extending \system.
P8 wanted to transfer the underlying concept into a visualization authoring context, where the encoding specifications are procedural and atomized---\iq{After analyzing the data, I need to present it with high-quality visualizations, but tools like ChatGPT often fail my expectations.}
P7 saw the value of a diagram in communication, especially to an audience with limited technical backgrounds. He said: \iq{I can use the scroll-telling in my presentation to explain how the data has been transformed.}
P3 envisioned a visual programming paradigm in which the basic building blocks can be self-composed or reused to communicate intention in addition to textual prompts to LLMs.

\section{Discussion}
In this section, we synthesize the implications and potential avenues for future research and reflect on the limitations.

\subsection{Design Implications}

\paragraph{Monitoring LLM agent through ``visible hands''}
Despite recent progress, known issues like hallucinations in LLM agents warrant external steering.
In \system, we abstract the LLM's generated content into high-level operations rather than raw text outputs, which align more closely with human cognitive processes.
Our approach also enriches the design space of \textit{AI resilient} interfaces~\cite{gu2024ai}.
Through static analysis, \system translates synthesized programs into abstracted operations.
These abstracted operations are brought to life through dynamic visual representations, making it possible for end-users to monitor the actions of LLM agents, similar to watching ``visible hands'' in real-time.
Future design may consider a similar mechanism of semantically rich representation and incremental update~\cite{chen2024sketch} in communicating agent actions.

\paragraph{Scrollytelling for LLM-generated content}
\system incorporates a basic form of scrollytelling, guiding users through the code by highlighting the corresponding diagrams as they scroll through the generated content. 
By combining the flow diagram with a scroll-triggered revealing mechanism, this technique aligns naturally with the generating process of LLM-produced content, fostering a deeper engagement and understanding of the content.
Looking ahead, we advocate developing automated streaming methods to create scrollytelling narratives for presenting LLM-generated content.
This complements the animation in the steaming generation phase, allowing users to control their understanding speed rather than passively following a predefined playing timeline.

\paragraph{\rev{Addressing context composition in different task granularity}}
\rev{One interesting property of LLMs is that they can provide reasonably high-quality responses to a wide variety of user tasks~\cite{subramonyam2024bridging}.
Echoing our formative study, users may request background information or incorporate more contexts when analyzing data.
They may start a sub-thread to test their assumptions~\cite{gu2024wizard}.
The highly diverse and evolving nature of user tasks in LLM-powered data analysis necessitates the development of adaptive user interfaces.
A more challenging direction is to generate visual representations for miscellaneous contexts in unpredictable LLM responses.}

\subsection{Future Works}

\paragraph{Democratizing data consumption with verifiable generative AI}
With nowadays generative AI, individuals without a programming background may easily create data visualizations for analysis or communication.
However, such democratization comes with challenges, particularly in ensuring the accuracy and reliability of AI-generated content.
There's a pressing need to navigate users to the potential inaccuracies and biases inherent in AI outputs~\cite{chen2024dashboard, kazemitabaar2024improving, zhutian2024generatingcodeevaluatinggpt}.
We believe that the key to fully leveraging AI's capabilities in data consumption hinges on creating user interfaces that align with the expertise levels of the intended users. 
In addition, different data tasks raise different requirements warranting tailored supports, such as an emphasis on the authorial intent matching of encoding schemes in expressive visualization design (\eg~\cite{vaithilingam2024dynavis,xie2023emordle}). 

\paragraph{\rev{Introducing a ``stop'' mechanism in human-LLM agent interaction}}
\rev{While \system is based on a chatbot-like interface, such an interaction paradigm can apply to a standalone AI assistant integrated into data analysis software or notebook platforms~\cite{mcnutt2023notebook}.
Essentially, during the ongoing conversation with LLM agents, users may be overwhelmed by the token-based output and fail to prevent propagating errors in time.
WaitGPT integrates proactive strategies to identify and rectify potential failures in AI-generated content.
Similarly, future works may further enrich the design space of visual representations of LLM outputs~\cite{gu2024ai, cai2024low} for instant understanding and explore a low-cost approach to facilitate steering content generation based on intermediate outputs.}

\paragraph{Exploiting interaction modalities in conversational data interface}
First, beyond textual prompts with simple selections of data slices in ChatGPT, future systems may incorporate other input types like direct manipulation~\cite{masson2023directgpt}, demonstration~\cite{huang2024table}, and reference~\cite{xie2023wakey}.
Second, to navigate users in nuanced decisions with drill-down explorations~\cite{gu2024how, gu2024wizard}, it is promising to provide explanations on demand~\cite{mehrpour2023tool}, or establish a tighter connection between code, data, textual analysis, and generated visualizations~\cite{wang2024wonderflow, dataparticles}.
Last, enabling users to directly reuse the generated code or interact with the resulting visualizations for further exploration~\cite{weng2024insightlens, gadhave2022reusing} could augment the flexibility of conversational data analysis tools.

\subsection{Limitation}
\paragraph{\rev{Threats to validity}}
The sample size in our formative and evaluation studies is relatively small and thus may not be representative of the broader population of data analysts and LLM users.
\rev{In the evaluation study, both conditions were equipped with standard syntax highlight for Python language.
However, without a careful visual design for key operations in the \baseline, participants may favor more on \system with its simplified information.} 
Besides, participants were prompted to view the transformable representation of the data analysis script, which may not reflect their natural interaction patterns. 
The reported usability rating may also be subject to response bias~\cite{dell2012yours} \rev{and participants' familiarity with the tasks}.
Future works may investigate how and how often users leverage this augmented view in their natural working space without explicit prompts to capture its real-world utility.

\paragraph{\rev{Scalability issues}}
In the framework, translating code into a flow diagram requires static analysis, which is dependent on the syntax.
\system is currently tailored to Python language and libraries like \texttt{Pandas} and \texttt{Matplotlib} for tubular data.
A potential solution to improve generalizability is to redesign LLM prompts to allow a mixed output stream of code and underlying operation objects, \eg~\cite{suh2023sensecape, kazemitabaar2024improving}.
However, the code stream visualization may not work for SQL-like languages with a reversed execution order compared to the procedure declaration.
Second, the flow diagram assumes a linear structure in the code, targeting fluent interfaces~\cite{shrestha2021unravel}.
Future works can incorporate control flows like loops and visual primitives for other data types.
\rev{Last, the current glyph design may not scale to tables with over 20 columns. To address this, unused columns can be aggregated, or important ones can be hidden.}

\section{Conclusion}
In this paper, we introduced \system, a novel interface design that transforms LLM-generated code into an accessible, interactive representation to address the reliability issues and user challenges in LLM-powered data analysis tools.
Drawing from an interview study with general users (N=8) of ChatGPT, we gained insights into general perspectives on these nascent tools and glitches in disruptive workflow, code verification, and labor-intensive iterations.
By translating stream-based code into a growing visualization of the key data operations and affording granular interactions, \system empowers users to monitor and steer data analysis performed by LLM agents.
A user study (N=12) covering basic data analysis tasks demonstrated that \system could enhance error detection rate and improve overall confidence in the results.

Our work contributes to the field of human-AI collaboration in data analysis by demonstrating the effectiveness of transformable code representations in facilitating user understanding and engagement.
As LLM applications in data analysis become more prevalent, prioritizing user experience and trust through accessible, interactive interfaces will be crucial in harnessing the potential of these powerful tools while ensuring their reliability and usability.
We urge more exploration of novel human-LLM interaction paradigms and intuitive visual representation design for LLM responses.

\begin{acks}
This research is supported by RGC GRF grant 16210321. The first author thanks Prof. Hanspeter Pfister for hosting the visit to the Harvard Visual Computing Group. We also thank the anonymous reviewers for their constructive feedback, the participants in the formative and user studies, and Zhan Wang, Leixian Shen, Xiaofu Jin, Shuchang Xu, Dr. Yanna Lin, Dr. Qingyu Guo, and Dr. Yun Wang for their valuable input.
\end{acks}
\newpage
\balance
\bibliographystyle{ACM-Reference-Format}
\bibliography{reference}

\appendix

\end{document}